\documentclass{article}

\usepackage{arxiv}

\usepackage[utf8]{inputenc} 
\usepackage[T1]{fontenc}    
\usepackage{hyperref}       
\usepackage{url}            
\usepackage{booktabs}       
\usepackage{amsfonts}       
\usepackage{nicefrac}       
\usepackage{microtype}      
\usepackage{lipsum}		
\usepackage{graphicx}
\usepackage{natbib}
\usepackage{doi}
\usepackage{amsmath} 
\usepackage{amssymb}
\usepackage{xcolor}

\title{Ab initio investigation on structural stability and phonon-mediated superconductivity in 2D-hydrogenated M$_2$X (M= Mo, V, Zr; X=C, N) MXene monolayer}

\author{ \href{https://orcid.org/0009-0004-2196-8245}{\includegraphics[scale=0.06]{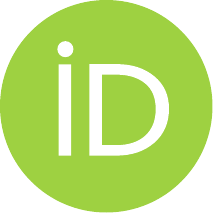}\hspace{1mm}Jakkapat Seeyangnok$^{*}$} \\
	Department of Physics\\
    Faculty of Science\\
	Chulalongkorn University\\
	Bangkok, Thailand \\
	\texttt{jakkapatjtp@gmail.com} \\
	\And
	\href{https://orcid.org/0000-0002-8450-7751}{\includegraphics[scale=0.06]{orcid.pdf}\hspace{1mm}Udomsilp Pinsook} \\
	Department of Physics\\
    Faculty of Science\\
	Chulalongkorn University\\
	Bangkok, Thailand \\
	\texttt{Udomsilp.P@Chula.ac.th} \\
}

\begin{document}
\maketitle

\begin{abstract}
We present a comprehensive first-principles study of hydrogenated M$_2$X (M = Mo, V, Zr; X = C, N) MXene monolayers, focusing on their structural stability, electronic properties, and superconducting behavior. Structural optimizations combined with phonon spectra reveal that partial hydrogenation (1H and 2H) is dynamically stable across most compositions, while full hydrogenation (4H) generally induces lattice instabilities. A notable exception is Zr$_2$CH$_4$, which retains dynamical stability even under maximum hydrogen coverage.  Electronic structure analysis shows that all hydrogenated MXenes remain metallic, with Fermi level dominated by transition-metal $d$ orbitals. In Zr$_2$CH$_4$, a Dirac-like band crossing at the Fermi level is observed, which is gapped by spin--orbit coupling (SOC), yielding a finite gap of $\sim$0.095~eV.  Electron–phonon coupling (EPC) calculations demonstrate that Mo-based MXenes exhibit strong EPC, with coupling constants $\lambda = 0.95$ (Mo$_2$CH), $1.23$ (Mo$_2$NH), and $1.55$ (Mo$_2$NH$_2$), corresponding to superconducting critical temperatures $T_c \approx 15$–22~K within the Allen--Dynes framework ($\mu^{*} = 0.10$). By contrast, V- and Zr-based MXenes display weak EPC and negligible $T_c$, with Zr$_2$CH$_4$ being a special case hosting Dirac-like states rather than superconductivity.  Our findings highlight hydrogen functionalization as an effective strategy to stabilize MXene monolayers and to tune their low-energy physics, revealing Mo-based nitride MXenes as promising phonon-mediated superconductors, while Zr$_2$C4H$_4$ emerges as a candidate for Dirac physics. 
\end{abstract}

\keywords{Superconductivity \and 2D Materials \and Hydrogenated MXene}

\section{Introduction}
Since the discovery of graphene in 2004~\cite{novoselov2004electric}, two-dimensional (2D) materials have become a cornerstone of condensed matter physics and nanotechnology, offering unconventional quantum phenomena and diverse technological applications enabled by their reduced dimensionality. At the same time, the search for high-temperature superconductors within the BCS framework~\cite{frohlich1950theory,migdal1958interaction,eliashberg1960interactions,nambu1960quasi} has highlighted the role of hydrogen, whose light mass gives rise to a high phonon spectrum that can significantly enhance electron–phonon coupling. This concept has been reinforced by intensive theoretical and experimental studies of high-pressure hydrides, motivated by Ashcroft’s pioneering predictions of metallic hydrogen and hydrogen-rich alloys~\cite{ashcroft1968metallic,ashcroft2004hydrogen}, which reshaped the pursuit of high-$T_c$ superconductivity. These two research frontiers converge in the exploration of hydrogen-rich 2D materials capable of hosting superconductivity under ambient or moderate pressures.

Several proposals exemplify this strategy, including doped and hydrogenated graphane~\cite{sofo2007graphane,savini2010first}, hydrogenated Ca-intercalated graphene bilayers~\cite{seeyangnok2025hydrogenation_nanoscale}, hydrogen-functionalized MgB$_2$~\cite{bekaert2019hydrogen}, and hydrogenated HPC$_3$~\cite{li2022phonon}, alongside subsequent studies of other hydrogenated 2D compounds such as transition-metal diborides~\cite{han2023high,seeyangnok2025high_npj2d} and LiBC monolayers~\cite{liu2024three}. All of these works underscore the promise of hydrogenation as a route to designing new superconductors.

Hydrogenation also introduces an additional degree of tunability by modifying both electronic and magnetic properties. For instance, Janus MoSH monolayers—synthesized via SEAR by replacing the top sulfur layer of MoS$_2$ with hydrogen—have been predicted to exhibit superconductivity with $T_c \approx 27$ K~\cite{lu2017janus,liu2022two,ku2023ab,pinsook2025superconductivity}. Related systems such as MoSLi~\cite{xie2024strong} and MoSeLi~\cite{moseliseeyangnok} display two-gap superconductivity, while dynamically stable WSH and WSeH monolayers are likewise predicted to superconduct with $T_c$ values exceeding 12 K~\cite{wseh_prb,wsh_2dmat,gan2024hydrogenation,fu2024superconductivity}. Moreover, Janus transition-metal chalcogenide hydrides (MXH, M = Ti, Zr, Hf; X = S, Se, Te) have been proposed as superconductors in the 10–30 K range~\cite{li2024machine,ul2024superconductivity}, though competing magnetic phases have also been reported~\cite{seeyangnok2025competition,sukserm2025half}, emphasizing their multifunctional character~\cite{yan2022enhanced,han2023theoretical,xue2024realization}.

Within this context, MXenes have emerged as particularly attractive candidates. Theoretical studies have examined the impact of hydrogenation on their superconducting properties. For example, hydrogenated Mo$_2$C is predicted to exhibit $T_c$ values up to 13 K~\cite{lei2017predicting}, while hydrogenated Ti-based MXenes~\cite{tsuppayakorn2023hydrogen}, including Ti$_2$CH, Ti$_2$CH$_2$, and Ti$_2$CH$_4$, show superconducting transitions at 0.2 K, 2.3 K, and 9.0 K, respectively. However, systematic investigations of hydrogenated MXenes beyond Ti-based systems remain scarce. Mo-based MXenes, in particular, have been studied for their rich physical properties, including superconductivity in Mo$_2$C and Mo$_2$N~\cite{bekaert2020first,pereira2022strain,liu2023theoretical}. The calculated superconducting transition temperatures ($T_c$) of the 2H and 1T phases are 7.1 K~\cite{bekaert2020first} and 3.2 K~\cite{lei2017predicting}, respectively, for Mo$_2$C, and 16.0 K~\cite{bekaert2020first} and 24.7 K~\cite{pereira2022strain} for Mo$_2$N. For V- and Zr-based MXenes, compounds such as V$_2$C, V$_2$N, Zr$_2$C, and Zr$_2$N have been investigated due to their wide range of attractive properties, including optical characteristics~\cite{lee2021investigation}, hydrogen storage potential~\cite{saharan2024v2n,yorulmaz2020systematical}, and suitability for energy storage in batteries~\cite{liu2022two,meng2018theoretical,behl2024recent}.

Motivated by these open questions, we present a comprehensive first-principles study of hydrogenated 1T-M$_2$X (M = Mo, V, Zr; X = C, N) MXene monolayers due to asymmetric properties. We show that hydrogen functionalization not only stabilizes their lattice structures but also reshapes their electronic spectra and significantly enhances electron–phonon coupling, resulting in phonon-mediated superconductivity with experimentally accessible $T_c$ values. These findings establish hydrogenated MXenes as a versatile platform for engineering superconductivity in two-dimensional materials.

\section{Methods}
First-principles calculations were performed within density functional theory (DFT) using \textsc{Quantum Espresso} (QE) \cite{giannozzi2009quantum, giannozzi2017advanced}. Structures were visualized in \textsc{VESTA} \cite{momma2011vesta} and optimized with the Broyden–Fletcher–Goldfarb–Shanno (BFGS) algorithm \cite{BFGS, liu1989limited} on a $24\times24\times1$ Monkhorst–Pack grid \cite{monkhorst1976special}. Relaxations employed a $0.02$ smearing width, a vacuum spacing of $20$ \AA, and force convergence below $10^{-5}$ eV/\AA. The exchange–correlation energy was described by the PBE functional \cite{perdew1996generalized} within the GGA, along with optimized norm-conserving Vanderbilt pseudopotentials \cite{hamann2013optimized, schlipf2015optimization}. Plane-wave cutoffs of 80 Ry (wavefunctions) and 320 Ry (charge densities) were used.

Phonon spectra and electron–phonon interactions were evaluated via density functional perturbation theory (DFPT). Phonons were calculated on a $24\times24\times1$ k-point grid and $8\times8\times1$ q-mesh to investigate phonon stability, while electron–phonon coupling used denser samplings  $24\times24\times1$ k-grids and $12\times12\times1$ q-grids. The superconducting transition temperature $T_{C}$ was estimated using the Allen–Dynes formula \cite{allen1975transition,pinsook2024analytic} with practical the Coulomb pseudopotential of $\mu^* = 0.1$.

To evaluate the thermodynamic stability of the hydrogenated M$_2$Y monolayers, we calculated the formation energy ($E_{\mathrm{form}}$) with respect to the pristine M$_2$Y host and molecular hydrogen (H$_2$). This approach reflects the realistic synthesis route, in which hydrogenation occurs through the incorporation of H$_2$ molecules. The formation energy is expressed as  
\begin{equation}
E_{\mathrm{form}} = (x+3)E_{\mathrm{M_2YH_x}} - \left( 3E_{\mathrm{M_2Y}} + x E_{\mathrm{H_2}} \right),
\end{equation}
where $E_{\mathrm{M_2YH_x}}$ (eV/atom) is the total energy of the hydrogenated system, 
$E_{\mathrm{M_2Y}}$ (eV/atom) is the energy of the pristine M$_2$Y monolayer, and 
$E_{\mathrm{H_2}}$ (eV/atom) represents the energy of an isolated hydrogen molecule. 
Here, $x$ denotes the number of hydrogen atoms adsorbed per formula unit of the M$_2$Y monolayer. 
A negative $E_{\mathrm{form}}$ (eV/cell) indicates that the hydrogenation process is exothermic and thermodynamically favorable, 
implying that the proposed monolayers could be experimentally realizable.

\section{Results and Discussion}
\subsection{Crystal Structure}
    \begin{figure}[h!]
		\centering
		\includegraphics[width=16cm]{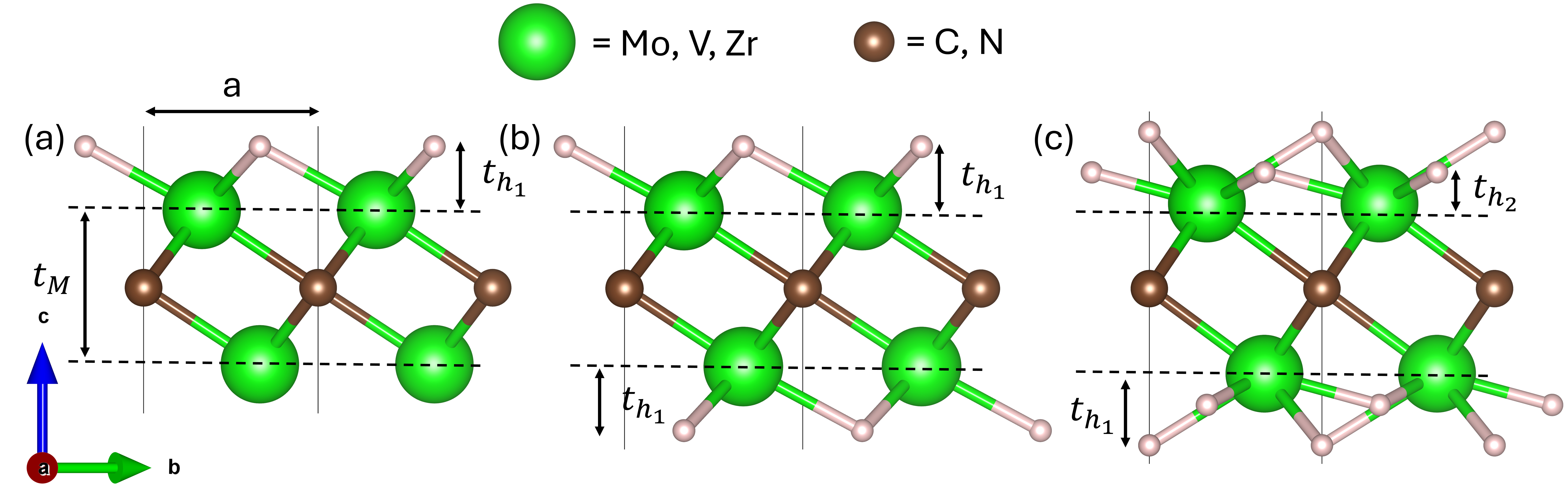}
		\caption{Optimized structures of hydrogenated M$_2$X (M = Mo, V, Zr; X = C, N) MXene monolayers. (a) Pristine M$_2$X monolayer with lattice parameter $a$ and interlayer distance between transition metals $t_M$. (b) Single-side hydrogenation showing hydrogen–metal distance $t_{h_1}$. (c) Double-side hydrogenation with hydrogen–metal distances $t_{h_1}$ and $t_{h_2}$. Green, brown, and pink spheres represent transition metals (Mo, V, Zr), C/N atoms, and hydrogen atoms, respectively.}
		\label{fig:structure}
	\end{figure}
    
To study the hydrogenation of phase of M$_2$X (M = Mo, V, Zr; X = C, M) MXene monolayers, we investigated possible hydrogenation of the trigonal space group \( P\overline{3}m1 \) (No.~156) M$_2$X monolayer, as illustrated in Figure~\ref{fig:structure}. The carbon atom forms the central layer, occupying the fractional coordinate (0,0) in the \textit{xy}-plane, while the transition metal atoms occupy the top and bottom layers at the fractional coordinates (2/3, 1/3) and (1/3, 2/3) in the \textit{xy}-plane, respectively. The distance between these two metal layers is denoted as $t_{M}$. Different concentrations of hydrogenation were considered, corresponding to one (1H), two (2H), three (3H) and four (4H) hydrogen atoms per unit cell. Consequently, the dynamical stability is investigated using positive-definite phonons. 

\begin{figure}[h!]
		\centering
		\includegraphics[width=12cm]{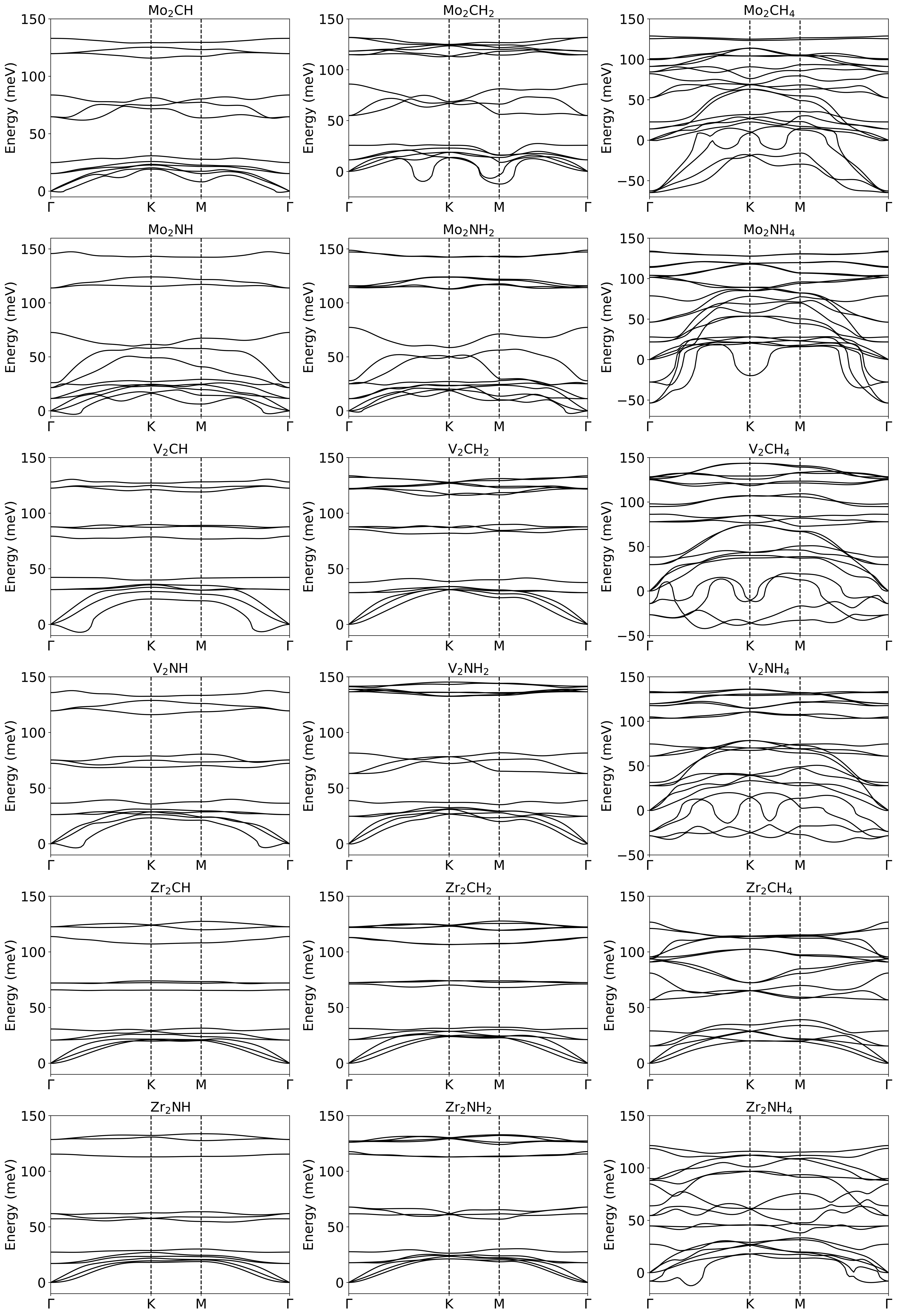}
		\caption{Phonon dispersion relations of hydrogenated M$_2$X (M = Mo, V, Zr; X = C, N) MXene monolayers with different hydrogen coverages: 1H (–CH or –NH), 2H (–CH$_2$ or –NH$_2$), and 4H (–CH$_4$ or –NH$_4$).  All MXenes are dynamically stable at 1H coverage, while most remain stable at 2H coverage except for Mo$_2$C. At full coverage (4H), imaginary frequencies appear in nearly all systems, indicating dynamical instability, with the exception of Zr$_2$C–CH$_4$, which retains phonon stability.}
		\label{fig:allphonons}
	\end{figure}

The structural parameters of hydrogenated M$_2$X MXene monolayers are summarized in Table~\ref{tab:structure}, including the lattice constant $a$, the interlayer transition-metal spacing $t_M$, and the hydrogen–metal distances $t_{h_1}$ and $t_{h_2}$. From Table~\ref{tab:structure}, The phonon stability of each configuration is also reported. From these results, it is evident that single- and double-hydrogenated structures are generally stable for most compositions, while full hydrogenation (4H) often leads to dynamical instabilities, as indicated by the presence of imaginary phonon modes as shown in Figure~\ref{fig:allphonons}. An exception is observed for Zr$_2$C, where even the fully hydrogenated configuration remains stable. This analysis highlights the critical role of hydrogen concentration in governing the structural and dynamical stability of MXene monolayers, suggesting that partial hydrogenation provides a more robust route for stabilizing these two-dimensional materials.

    \begin{table}[h!]\label{tab:structure}
    \centering
    \caption{Structural parameters of hydrogenated M$_2$X (M = Mo, V, Zr; X = C, N) MXene monolayers: lattice constant $a$ (\AA), interlayer distance between transition metals $t_M$ (\AA), hydrogen–metal distances $t_{h_1}$ and $t_{h_2}$ (\AA), formation energy (eV/cell) and phonon stability.}
	   \begin{tabular}{|c|c|c|c|c|c|c|c|}
        \hline
    MXene & hydrogenation & Lattice $a$ & $t_M$ & $t_{h_1}$ ($t_{h_2}$) & E$_{\textbf{form}}$ (eV)& phonon stability \\
	   \hline
	Mo$_2$C & 1H & 2.93 & 2.51 & 1.04 & -0.77 & stable\\
            & 2H & 2.93 & 2.56 & 1.05 & -1.48 & unstable \\
            & 4H & 2.96 & 2.56 & 1.16 (1.11) & -0.34 & unstable \\
            \hline
    Mo$_2$N & 1H & 2.80 & 2.84 & 1.15 & -0.87 & stable\\
            & 2H & 2.79 & 2.88 & 1.15 & -1.75 & stable\\
            & 4H & 2.87 & 2.78 & 1.20 (1.10) & -0.71 & unstable \\
            \hline
    V$_2$C & 1H & 2.91 & 2.14 & 0.97 & -1.08 & stable\\
           & 2H & 2.88 & 2.23 & 0.99 & -2.03 & stable\\
           & 4H & 2.85 & 2.39 & 1.17 (0.90)& -0.85 & unstable \\
           \hline
    V$_2$N & 1H & 2.86 & 2.18 & 1.00 & -0.92 & stable\\
           & 2H & 2.88 & 2.23 & 0.99 & -1.92 & stable\\
           & 4H & 2.83 & 2.37 & 1.15 (0.96) & -0.76 & unstable \\
            \hline
    Zr$_2$C & 1H & 3.28 & 2.52 & 1.04 & -1.39 & stable \\
            & 2H & 3.30 & 2.50 & 1.03 & -2.79 & stable \\
            & 4H & 3.27 & 2.80 & 1.18 (0.51) & -2.77 & stable \\
            \hline
    Zr$_2$N & 1H & 3.25 & 2.47 & 1.02 & -1.39 & stable \\
            & 2H & 3.24 & 2.49 & 1.04 & -2.73 & stable \\
            & 4H & 3.24 & 2.75 & 1.18 (0.61) & -2.65 & unstable \\
            \hline
	   \end{tabular}
	\end{table}

The formation energies ($E_{\text{form}}$) as shown in Table~\ref{tab:structure} of hydrogenated M$_2$X MXenes are consistently negative, confirming the thermodynamic favorability of hydrogen adsorption. At single hydrogen coverage (1H), all systems are both energetically favorable and dynamically stable. Increasing the coverage to two hydrogens per unit cell (2H) further reduces $E_{\text{form}}$, while most MXenes remain phonon-stable. Under full coverage (4H), however, phonon instabilities appear in nearly all cases despite negative formation energies, with Zr$_2$C being the only MXene that retains stability at this concentration as shown in Figure~\ref{fig:allphonons}.

The phonon dispersion relations~as shown in Figure~\ref{fig:allphonons} provide deeper insight into this trend. At low coverage, the spectra exhibit well-defined acoustic (ZA,TA,LA) and optical branches without imaginary frequencies, confirming dynamical stability. At 2H coverage, most systems maintain positive phonon modes, although soft modes emerge in some cases, such as Mo$_2$CH at M point, signaling a tendency toward structural instability in Mo$_2$CH$_2$. Under full coverage (4H), pronounced imaginary frequencies appear along multiple high-symmetry directions, revealing strong dynamical instabilities. The sole exception is Zr$_2$CH$_4$, which maintains a positive phonon spectrum across the Brillouin zone and thus remains stable even under maximum hydrogenation. These results highlight that while hydrogen adsorption is energetically favorable, excessive hydrogen concentration generally destabilizes the lattice, with Zr$_2$C standing out as the most robust host for high hydrogen coverage.

    \begin{figure}[h!]
		\centering
		\includegraphics[width=8cm]{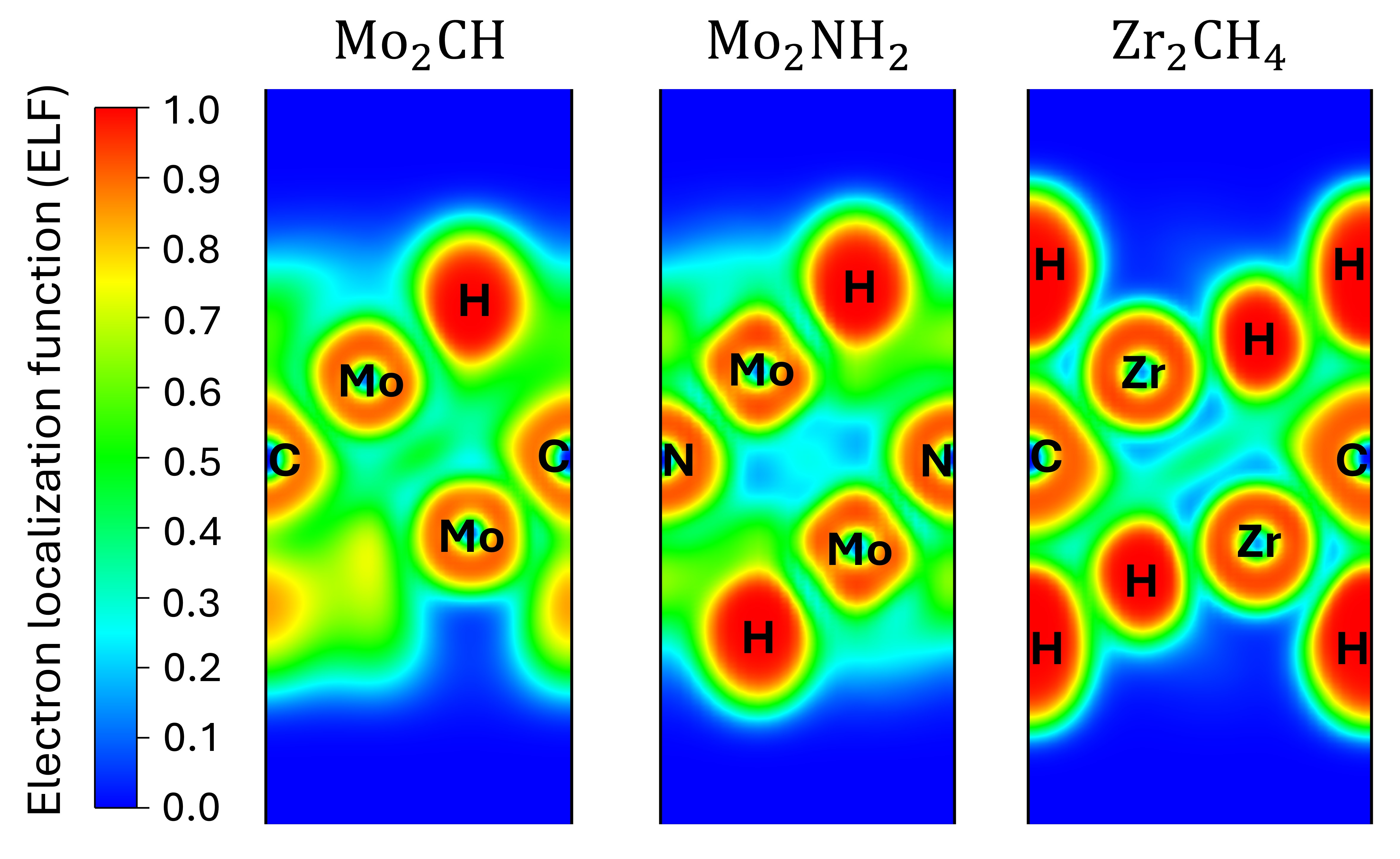}
		\caption{Electron localization function (ELF) maps of hydrogenated MXenes: Mo$_2$CH, Mo$_2$NH$_2$, and Zr$_2$CH$_4$. The strong localization around hydrogen atoms indicates concentrated electron density, whereas the absence of attractors between transition metals (Mo, Zr) and neighboring atoms (H, C, N) reflects metallic bonding.}

		\label{fig:elf}
	\end{figure}

The electron localization function (ELF) provides valuable insight into the bonding nature of hydrogenated M$_2$X MXenes. As shown in Figure~\ref{fig:elf}, pronounced electron localization is observed around the hydrogen atoms, indicating that electrons are strongly concentrated in their vicinity and form well-defined H-related states. In contrast, no significant ELF attractors are found between the transition metal (Mo or Zr) and hydrogen, or between the transition metal and carbon atoms, which is consistent with the metallic bonding character of MXenes.

\subsection{Electronic Structure}
    \begin{figure}[h!]
		\centering
		\includegraphics[width=12cm]{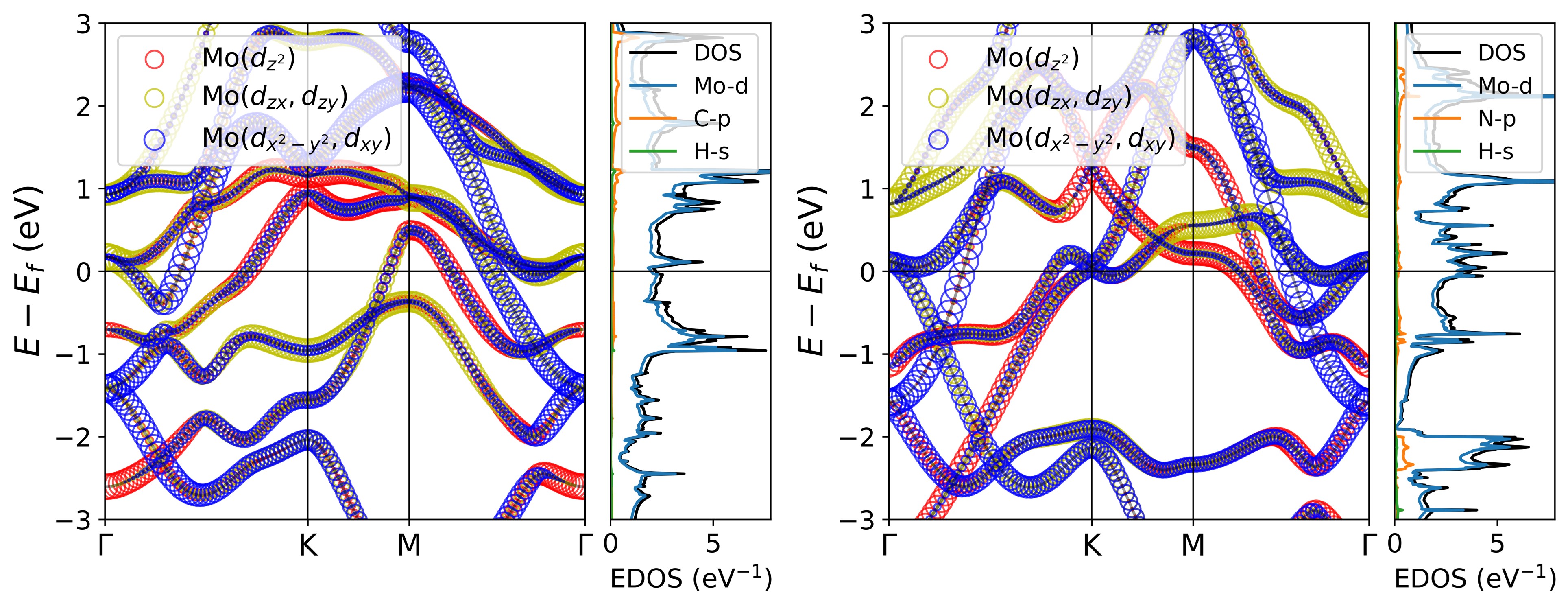}
		\caption{Orbital-projected electronic band structures and corresponding density of states (DOS) of hydrogenated Mo$_2$X MXenes: (a) Mo$_2$CH and (b) Mo$_2$NH$_2$. The band structures are projected onto Mo-\(d\) orbitals, highlighting contributions from \(d_{z^2}\) (red), \(d_{zx}, d_{zy}\) (yellow), and \(d_{x^2-y^2}, d_{xy}\) (blue). The DOS panels show the total DOS (black) along with orbital contributions from Mo-\(d\), C-\(p\)/N-\(p\), and H-\(s\) states. The results reveal that the electronic states near the Fermi level are dominated by Mo-\(d\) orbitals, with additional hybridization from C/N-\(p\) and H-\(s\) states.}
		\label{fig:bands_dos}
	\end{figure}

The electronic band structures and density of states (DOS) of hydrogenated Mo$_2$X MXenes are presented in Figure~\ref{fig:bands_dos}, where the orbital-projected contributions highlight the role of Mo $d$ orbitals in governing the electronic properties. For Mo$_2$CH as shown in Figure~\ref{fig:bands_dos}~(a), the states near the Fermi level ($E_f$) are primarily composed of Mo $d_{z^2}$, $d_{zx}$/$d_{zy}$, and $d_{x^2-y^2}$/$d_{xy}$ orbitals. A similar trend is observed for Mo$_2$NH$_2$ as shown in Figure~\ref{fig:bands_dos}(b), where the Mo $d$ orbitals dominate the bands close to $E_f$, while the N-$p$ and H-$s$ orbitals contribute mainly at deeper energy levels. The DOS profiles further confirm these observations, showing sharp peaks around the Fermi level that are dominated by Mo $d$ states. The contributions from C-$p$ (in Mo$_2$CH) and N-$p$ (in Mo$_2$NH$_2$) orbitals are less pronounced but still important for stabilizing the bonding environment and modulating the band dispersion.

For other hydrogenated M$_2$X MXenes studied in this work (V$_2$X and Zr$_2$X), the orbital characteristics remain qualitatively similar: the transition metal $d$ orbitals dominate the vicinity of the Fermi level across all hydrogenation concentrations. The $p$ orbitals of C or N atoms hybridize with the metal $d$ states at lower energies, while H-$s$ orbitals introduce additional states but do not significantly alter the electronic metallicity. These results demonstrate that the physics of hydrogenated M$_2$X MXenes is universally governed by the transition metal $d$ orbitals, while surface terminations and hydrogenation primarily tune the degree of hybridization and band dispersion rather than fundamentally changing the orbital character near $E_f$.

    \begin{figure}[h!]
		\centering
		\includegraphics[width=12cm]{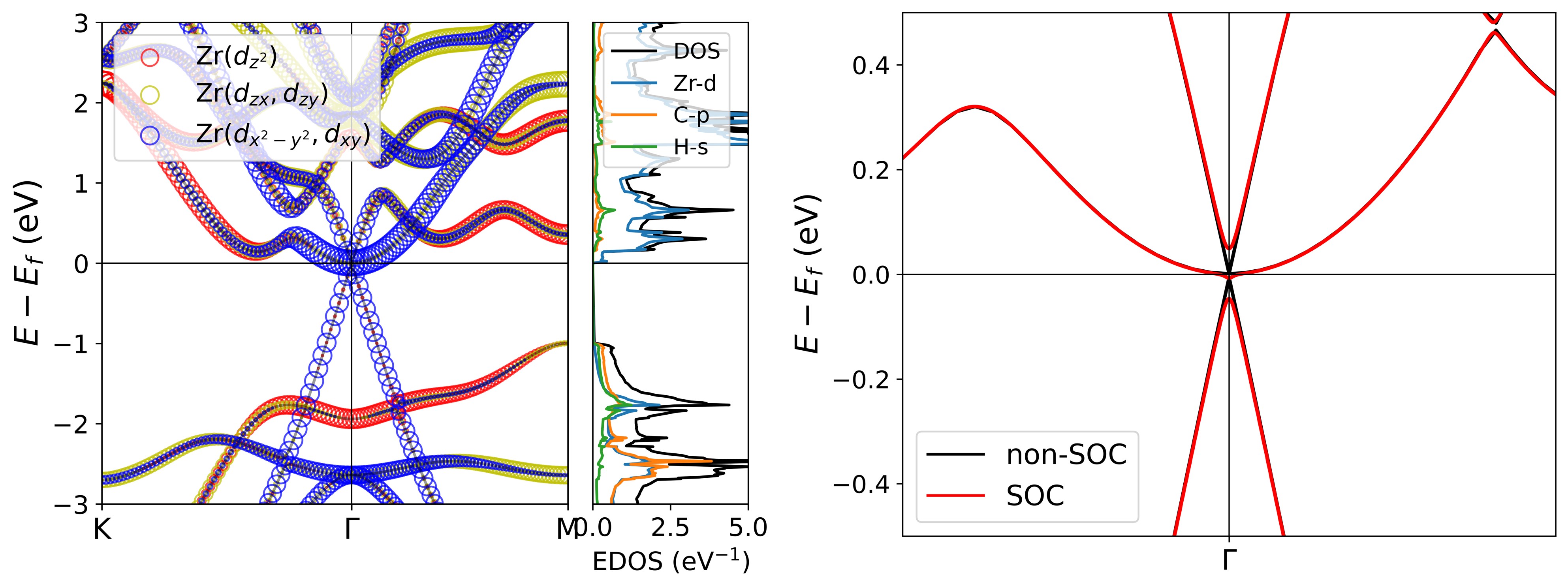}
		\caption{(Left) Orbital-projected electronic band structure and density of states (DOS) of hydrogenated Zr$_2$CH$_4$ without spin--orbit coupling (SOC). The low-energy states around the Fermi level are dominated by Zr $d$ orbitals, with additional contributions from C-$p$ and H-$s$ states. (Right) Comparison of the electronic band dispersion near the $\Gamma$ point with and without SOC. The inclusion of SOC lifts the degeneracy at the Dirac-like crossing, opening a small gap and slightly modifying the band curvature.}
    \label{fig:zr2ch4_soc}
	\end{figure}

To further examine the role of relativistic effects, we investigated the electronic band structure of hydrogenated Zr$_2$CH$_4$ with and without spin--orbit coupling (SOC). As shown in Fig.~\ref{fig:zr2ch4_soc}, the non-SOC calculation reveals a Dirac-like band crossing at the $\Gamma$ point, arising primarily from the Zr $d$ orbitals. This behavior was also theoretically predicted in the Zr$_2$CCl$_2$ monolayer. Upon inclusion of SOC, this degeneracy is lifted, and a finite bandgap of approximately 0.095~eV opens at the Fermi level. The orbital-projected analysis confirms that the electronic states are dominated by Zr $d$ orbitals, while contributions from C-$p$ and H-$s$ states are relatively minor. The small SOC-induced gap in Zr$_2$CH$_4$ implies that its electronic structure is sensitive to external perturbations such as strain or electrostatic gating. Such tunability may drive the system into a topologically nontrivial phase, characterized by bulk insulating behavior with symmetry-protected conducting edge states, or alternatively enhance anisotropic charge transport properties.

\subsection{Phonon Structure} 
    \begin{figure}[h!]
	\centering
	\includegraphics[width=15cm]{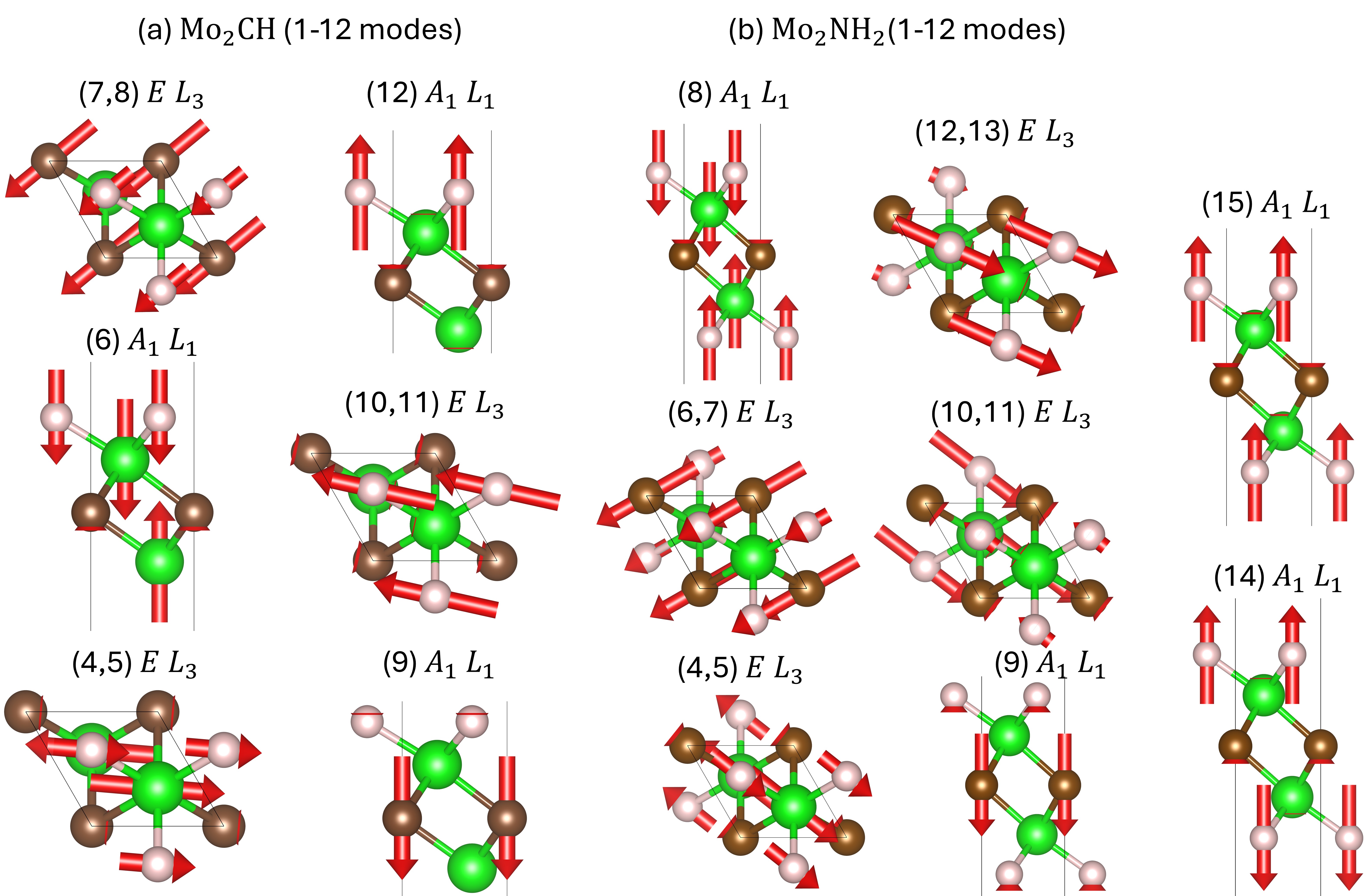}
	\caption{Phonon vibrational modes of functionalized Mo$_2$-based MXenes: 
    (a) Mo$_2$CH and (b) Mo$_2$NH$_2$. The numbers label the corresponding 
    eigenmodes for each system. Spheres of different colors denote Mo, C/N, and H atoms, 
    while red arrows indicate the direction and relative amplitude of atomic displacements.}
    \label{fig:phonon_modes}
	\end{figure}
For the phonon analysis, we examined the phonon spectra of hydrogenated Mo$_2$-based MXenes (Mo$_2$CH, Mo$_2$NH, and Mo$_2$NH$_2$), together with their corresponding phonon eigenmodes, summarized in Table~\ref{tab:phonon_comparison} and visualized in Figure~\ref{fig:phonon_modes}, motivated by their potential to exhibit promising superconducting properties. Since the electronic states near the Fermi level are dominated by Mo $d$ orbitals, we expect that the phonon modes responsible for electron--phonon coupling mainly arise from transition-metal vibrations.  

\begin{table}[h!]
\centering
\caption{Comparison of selected phonon vibrational modes (in meV) 
and their symmetry representations for Mo$_2$CH, Mo$_2$NH, and Mo$_2$NH$_2$. 
All listed modes are infrared- and Raman-active (I+R).}
\label{tab:phonon_comparison}
\begin{tabular}{|c|c|c|}
\hline
Mo$_2$CH & Mo$_2$NH & Mo$_2$NH$_2$ \\
\hline
15.4 \; (E L$_3$)   & 11.2 \; (E L$_3$)   & 11.3 \; (E L$_3$) \\
24.8 \; (A$_1$ L$_1$) & 19.8 \; (E L$_3$) & 23.3 \; (E L$_3$) \\
64.7 \; (E L$_3$)   & 26.2 \; (A$_1$ L$_1$) & 27.9 \; (A$_1$, L$_1$) \\
83.8 \; (A$_1$ L$_1$) & 71.5 \; (A$_1$ L$_1$) & 77.8 \; (A$_1$ L$_1$) \\
119.7 \; (E L$_3$)   & 115.1 \; (E L$_3$)   & 114.8 \; (E L$_3$) \\
132.9 \; (A$_1$ L$_1$) & 147.2 \; (A$_1$ L$_1$) & 115.3 \; (E L$_3$) \\
--                    & --                    & 147.2 \; (A$_1$ L$_1$) \\
--                    & --                    & 148.0 \; (A$_1$ L$_1$) \\
\hline
\end{tabular}
\end{table}

For the acoustic modes, the projected phonon density of states indicates that the majority of vibrations are associated with the transition metal atoms. For the optical modes, the symmetries follow the irreducible representations of the point group, consisting of doubly degenerate E (L$_3$) modes and non-degenerate A$_1$ (L$_1$) modes. Both types are infrared- and Raman-active, reflecting strong optical responses. In all three systems, the low-frequency modes are dominated by Mo in-plane vibrations, whereas the high-frequency branches are mainly associated with light-element vibrations (C, N, and H).  

For Mo$_2$CH and Mo$_2$NH, at the $\Gamma$ point, the optical phonon spectrum consists of three in-plane vibration modes and three out-of-plane vibration modes ($\Gamma = 3E + 3A_1$). In Mo$_2$CH, the transition-metal in-plane vibration (E L$_3$) mode appears at 15.4~meV, followed by the transition-metal out-of-plane vibration (A$_1$ L$_1$) mode at 24.8~meV. In Mo$_2$NH, the transition-metal in-plane vibration (E L$_3$) mode shifts lower to 11.4~meV, while the transition-metal out-of-plane vibration (A$_1$ L$_1$) mode shifts higher to 26.2~meV. This suggests that higher-energy phonon modes are more strongly coupled to the electrons, potentially enhancing electron--phonon coupling.  

For the case of higher hydrogen concentration in Mo$_2$NH$_2$, the optical phonons associated with transition-metal vibrations are shifted to even higher energies, with the transition-metal out-of-plane mode (A$_1$ L$_1$) appearing at 27.9~meV. This further increase in phonon energy may lead to stronger electron--phonon coupling.

\subsection{Phonon-Mediated Superconductivity}
    \begin{figure}[h!]
	\centering
	\includegraphics[width=17cm]{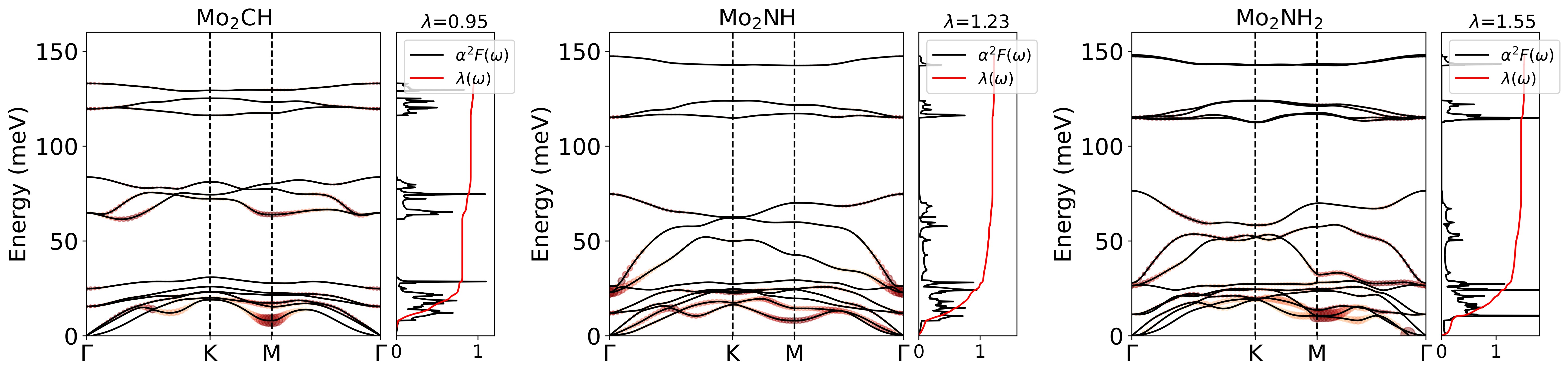}
	\caption{EPC--weighted phonon dispersion relations of hydrogenated Mo$_2$X MXenes: 
    (a) Mo$_2$CH, (b) Mo$_2$NH, and (c) Mo$_2$NH$_2$. 
    The size of the red dots represents the mode-resolved electron--phonon coupling strength, 
    $\lambda_{q\nu}$. Right panels show the Eliashberg spectral function, $\alpha^{2}F(\omega)$ (black curves), 
    and the cumulative electron--phonon coupling constant, $\lambda(\omega)$ (red curves). 
    The total EPC constants are found to be $\lambda = 0.95$, $1.23$, and $1.55$, respectively. 
    The results indicate that low- and mid-frequency phonon modes dominate the electron--phonon interaction, 
    with nitrogen functionalization significantly enhancing the overall coupling.}
    \label{fig:phonon_lambda}
	\end{figure}

The superconducting properties of hydrogenated Mo$2$X MXenes were investigated through an analysis of the electron–phonon interaction. Figure~\ref{fig:phonon_lambda} shows the phonon dispersion relations weighted by the electron–phonon coupling (EPC) strength, $\lambda{q\nu}$, along with the Eliashberg spectral function $\alpha^{2}F(\omega)$ and the cumulative coupling constant $\lambda(\omega)$ for Mo$_2$CH, Mo$_2$NH, and Mo$_2$NH$2$. The EPC-weighted dispersions indicate that the dominant contributions to the coupling originate from low- and mid-frequency phonon branches. Since the electronic states near the Fermi level are dominated by Mo $d$ orbitals, these phonons are mainly associated with transition-metal vibrations. The mode-resolved coupling strength $\lambda{q\nu}$ (represented by the red dot size) is particularly pronounced in the acoustic and low-energy optical branches near the $\Gamma$ and M points, reflecting strong scattering channels mediated by lattice vibrations.

The corresponding $\alpha^{2}F(\omega)$ spectra show a clear evolution with functionalization. For Mo$_2$CH, the total EPC constant is calculated as $\lambda = 0.95$, which is already sufficient to suggest moderate electron–phonon-mediated superconductivity. Upon nitrogen substitution, $\lambda$ increases significantly to $1.23$ in Mo$_2$NH, and further to $1.55$ in Mo$_2$NH$_2$, demonstrating a strong enhancement of electron–phonon coupling strength. The cumulative $\lambda(\omega)$ curves indicate that most of the contribution arises from phonon modes below $\sim$80 meV, confirming that low-energy vibrations dominate the pairing mechanism.

    \begin{table}[h!]
    \centering
    \label{tab:superconducting}
    \caption{Superconducting parameters of hydrogenated M$2$X (M = Mo, V, Zr; X = C, N) MXene monolayers, including the electron–phonon coupling constant ($\lambda$), logarithmic average phonon frequency ($\omega{\text{log}}$), second moment of the phonon spectrum ($\omega_{2}$), and the superconducting critical temperature estimated from the Allen–Dynes formula ($T_C^{AD}$). A dash (–) indicates cases where superconducting properties are not applicable: specifically, Zr$_2$C with 4H hydrogenation exhibits a Dirac-cone-like band crossing at the Fermi level rather than a conventional superconducting instability.}
	   \begin{tabular}{lclclclclcl}
        \hline
    MXene & hydrogenation & $\lambda$ & $\omega_{\textbf{log}}$ (K) & $\omega_{2}$ (K) & T$_C^{AD}$ (K) \\
	   \hline
	Mo$_2$C & 1H & 0.95 & 220.9 & 425.9 & 15.1 \\
    Mo$_2$N & 1H & 1.23 & 177.2 & 342.6 & 18.2 \\
            & 2H & 1.55 & 159.4 & 388.0 & 21.7 \\
            \hline
    V$_2$C & 1H & 0.26 & 434.0 & 603.5 & 0.0 \\
           & 2H & 0.40 & 421.0 & 667.1 & 1.9 \\
    V$_2$N & 1H & 0.50 & 366.5 & 533.2 & 4.4 \\
           & 2H & 0.40 & 455.1 & 641.8 & 2.0 \\
            \hline
    Zr$_2$C & 1H & 0.22 & 261.1 & 402.5 & 0.0 \\
            & 2H & 0.11 & 687.9 & 718.1 & 0.0 \\
            & 4H & - & - & - & - \\
    Zr$_2$N & 1H & 0.36 & 273.1 & 423.1 & 0.7 \\
            & 2H & 0.51 & 321.0 & 504.9 & 4.1 \\
            \hline
	   \end{tabular}
	\end{table}
    
The superconducting parameters of hydrogenated M$_2$X (M = Mo, V, Zr; X = C, N) monolayers are summarized in Table~\ref{tab:superconducting}. Our results reveal a strong dependence of the electron--phonon coupling (EPC) strength and superconducting transition temperature ($T_c$) on both the transition metal and the hydrogenation level.  

For Mo-based systems, hydrogenation induces substantial EPC, with coupling constants $\lambda$ ranging from 0.95 in Mo$_2$CH to 1.55 in Mo$_2$N with double hydrogenation. The logarithmic average phonon frequencies ($\omega_{\text{log}}$) are moderately reduced upon increasing hydrogen content, reflecting phonon softening that enhances EPC. Consequently, the estimated $T_c^{AD}$ values span 15.1--21.7~K, placing Mo-based MXenes among the most promising 2D superconductors in this family.  

In contrast, V-based compounds exhibit considerably weaker EPC. For example, V$_2$CH yields $\lambda = 0.26$ and essentially vanishing $T_c^{AD}$, while further hydrogenation to V$_2$C2H only marginally increases $T_c^{AD}$ to 1.9~K. Nitrides show slightly stronger coupling, with V$_2$NH and V$_2$N2H giving $\lambda = 0.50$ and 0.40, corresponding to $T_c^{AD}$ of 4.4~K and 2.0~K, respectively. These results suggest that, although V-based MXenes are metallic and dynamically stable, their EPC is insufficient to support robust superconductivity under ambient conditions.  

For Zr-based MXenes, superconductivity is essentially suppressed. Zr$_2$CH exhibits a very weak EPC constant of $\lambda = 0.22$, leading to negligible $T_c^{AD}$. Increasing hydrogen coverage further reduces the EPC strength, with Zr$_2$C2H yielding $\lambda = 0.11$, while Zr$_2$C4H shows no conventional superconducting instability due to the presence of a Dirac-cone-like band crossing near the Fermi level. A small enhancement is observed in the nitride series: Zr$_2$NH and Zr$_2$N2H yield $T_c^{AD}$ values of 0.7~K and 4.1~K, respectively, but these remain much lower than their Mo counterparts.  

A noteworthy exception arises for Zr$_2$CH$_4$. Although dynamically stable, its electronic structure is characterized by a Dirac-cone-like band crossing at the Fermi level, rather than a conventional metallic density of states favorable for superconductivity. This behavior was also theoretically predicted in the Zr$_2$CCl$_2$ monolayer \cite{duan2022dirac}. Consequently, superconducting parameters such as $\lambda$ and $T_c$ are not applicable in this case, as indicated by the dashes in Table~\ref{tab:superconducting}. The emergence of Dirac-like features suggests that Zr$_2$CH$_4$ may instead be relevant for exploring massless fermions and topological transport phenomena, positioning it as a candidate material for Dirac physics rather than phonon-mediated superconductivity.  

Overall, these results highlight the strong chemical dependence of superconductivity in hydrogenated MXene monolayers. Mo-based hydrides emerge as the most promising candidates for 2D superconductivity with transition temperatures exceeding 20~K, while V- and Zr-based systems display either weak or negligible superconductivity. This contrast can be attributed to the dominant $d$-orbital character at the Fermi level in Mo-based compounds, which promotes strong EPC, whereas the electronic structures of V- and Zr-based systems yield weaker coupling. These findings suggest that careful choice of the transition metal and hydrogen coverage is critical for designing experimentally feasible superconducting 2D MXenes.  

\section{General Discussion}
The comparative analysis of 2D superconductors, including MXenes and other layered BCS-type materials, reveals several trends in electron-phonon coupling and critical temperatures. For the MXene systems studied, hydrogenation significantly influences both the electron-phonon coupling constant ($\lambda$) and the superconducting critical temperature ($T_c$). For instance, Mo$_2$N exhibits an increase in $\lambda$ from 1.23 in the 1H configuration to 1.55 in the 2H configuration, corresponding to a rise in $T_c^{AD}$ from 18.2~K to 21.7~K. This demonstrates that hydrogenation can effectively enhance the superconducting properties by modulating the phonon spectra, as reflected in changes in both the logarithmic average frequency ($\omega_{\text{log}}$) and the second moment frequency ($\omega_2$).

When compared to other 2D BCS superconductors, the MXenes show moderately strong coupling, with $\lambda$ values approaching or exceeding unity, similar to recently reported hydride-intercalated borides. For example, monolayer h-MgB$_2$ exhibits $\lambda = 1.46$ with $T_c \approx 67$~K \cite{bekaert2019hydrogen}, while Ti$_2$B$_2$H$_4$ has $\lambda = 1.18$ and $T_c \approx 48.6$~K \cite{han2023high}. Certain boride-based systems, such as V$_2$B$_2$H$_4$ ($\lambda = 1.29$, $T_c = 83$~K) and Nb$_2$B$_2$H$_4$ ($\lambda = 1.34$, $T_c = 69$~K) \cite{seeyangnok2025high_npj2d}, also exhibit strong coupling, demonstrating that both robust electron-phonon interaction and optimized phonon modes are critical for high-temperature superconductivity in two-dimensional materials.

In contrast, bulk and monolayer MgB$_2$ show relatively weaker coupling ($\lambda = 0.61$–0.68) with correspondingly lower $T_c$ \cite{nagamatsu2001superconductivity,bekaert2017evolution,cheng2018fabrication}, although chemical functionalization, such as hydrogenation in h-MgB$_2$, significantly enhances both $\lambda$ and $T_c$ \cite{bekaert2019hydrogen}. Similar trends are observed for LiCB systems, where hydrogenation increases $\lambda$ from 0.59 to 1.36 and enhances $T_c$ from 70~K to 80~K \cite{modak2021prediction,liu2024three}.

\section{Conclusion}
In this work, we systematically investigated the structural stability, electronic properties, and superconducting behavior of hydrogenated M$_2$X (M = Mo, V, Zr; X = C, N) MXene monolayers using first-principles calculations combined with electron–phonon coupling (EPC) theory. Our results show that hydrogenation provides an effective route to tune both the dynamical stability and the electronic properties of MXenes. From the structural analysis, we found that single- and double-side hydrogenation (1H and 2H) are generally stable across most compositions, whereas full hydrogenation (4H) often induces dynamical instabilities due to soft phonon modes. An important exception is Zr$_2$CH$_4$, which remains stable even under maximum hydrogenation.

Electronic structure calculations demonstrate that all studied hydrogenated MXenes retain their metallic character, with states near the Fermi level dominated by transition-metal $d$ orbitals. For Zr$_2$C–CH$_4$, relativistic effects introduce a finite SOC-induced band gap at the Dirac-like crossing near the Fermi level, suggesting opportunities to explore topological transport phenomena in addition to conventional metallic behavior. 

Phonon dispersion and EPC calculations reveal a strong dependence of superconducting properties on chemical composition and hydrogen coverage. In Mo-based MXenes, the EPC constants reach $\lambda = 0.95$ (Mo$_2$CH), $1.23$ (Mo$_2$NH), and $1.55$ (Mo$_2$NH$_2$), leading to predicted superconducting transition temperatures of $T_c \approx 15$–22~K within the Allen–Dynes framework ($\mu^* = 0.10$). The dominant contributions arise from low- and mid-frequency phonon modes below $\sim$80~meV, confirming their role as the primary pairing channels associated with $d$-orbital electronic states at the Fermi level and transition-metal vibrations. By contrast, V- and Zr-based MXenes generally exhibit weak EPC ($\lambda < 0.5$), resulting in negligible or very low $T_c$. The special case of Zr$_2$C–4H, where Dirac-like electronic states emerge in place of conventional superconductivity, further highlights the diverse phenomena accessible in hydrogenated MXenes.

Overall, our study demonstrates that hydrogen functionalization provides a powerful strategy to stabilize MXene monolayers and to engineer their superconducting properties. In particular, nitrogen-terminated Mo-based MXenes emerge as promising candidates for phonon-mediated superconductivity with experimentally accessible $T_c$ values. Meanwhile, the unusual Dirac-like features in Zr$_2$C–4H suggest a distinct avenue toward topological electronic phases. These insights establish hydrogenated MXenes as a versatile platform for exploring the interplay between structure, bonding, and correlated quantum phenomena in two-dimensional materials.
    \section*{Data Availability}
    The data that support the findings of this study are available from the corresponding
    authors upon reasonable request.
    
    \section*{Code Availability}
    The first-principles DFT calculations were performed using the open-source Quantum ESPRESSO package, available at \url{https://www.quantum-espresso.org}, along with pseudopotentials from the Quantum ESPRESSO pseudopotential library at \url{https://pseudopotentials.quantum-espresso.org/}. 

    \section*{Acknowledgements}
	This research project is supported by the Second Century Fund (C2F), Chulalongkorn University. We acknowledge the supporting computing infrastructure provided by NSTDA, CU, CUAASC, NSRF via PMUB [B05F650021, B37G660013] (Thailand). URL:www.e-science.in.th.

    \section*{Author Contributions}
    Jakkapat Seeyangnok performed all of the calculations, analysed the results, wrote the first draft manuscript, and coordinated the project. Udomsilp Pinsook analysed the results and wrote the manuscript.

    \section*{Conflict  of Interests}
    The authors declare no competing financial or non-financial interests.

	\section*{References}
    \bibliographystyle{unsrt} 
	\bibliography{references}

\end{document}